\documentstyle[preprint,aps,eqsecnum]{revtex}

\begin{document}

\def\ran#1{\langle#1\rangle}
\def\t{\displaystyle}
\def\cc{\c c\~ao }
\def\ccs{\c c\~oes }
\def\ii{\'\i }
\def\bra{\langle}
\def\ket{\rangle}
\def\psibar{\overline{\psi}}
\def\sen{{\rm \>sin\>}}
\def\senh{{\rm \>sinh\>}}
\def\per{1\!\!\!\!\perp}

\title{\bf Thermal Decays in a Hot Fermi Gas}

\author{\bf E. S. Fraga$^{\dag}$ and C. A. A. de Carvalho$^{\ddag}$}

\address{Instituto de F\'\i sica \\ Universidade Federal
do Rio de Janeiro \\ C.P. 68528, Rio de Janeiro, RJ, 21945-970, Brasil}

\date{}

\maketitle

\begin{abstract}
We present a study of the decay of metastable states of a scalar field via
thermal activation, in the presence of a finite density of fermions. The
 process we consider is the nucleation of ``{\it
droplets}'' of true vacuum inside the false one. We analyze a one-dimensional
 system of interacting bosons and fermions,
considering the latter at finite temperature and with a given chemical
 potential.
As a consequence of a non-equilibrium formalism previously developed, we
 obtain time-dependent decay
rates.
\end{abstract}

\section{Introduction}

The dynamics of metastable systems is a subject that may be found in all realms
 of Physics. In Condensed Matter Physics, in polymer transitions \cite{fraga}
 and liquid mixtures \cite{langer,Notes}; in High Energy Physics, in the
problem
 of baryon number violation \cite{particulas} and in the formation of chiral
 condensates in heavy ions collisions \cite{ions} ; in Astrophysics, in
``nucleating planets''; in Cosmology, in the well known Inflationary Models
 \cite{inflacao}. All these examples have at least one thing in common: the
 occurrence of metastability, which requires non-equilibrium methods.

As a model for learning something about the qualitative features of the decays
 of metastable states, we will study a one-dimensional system of interacting
fermions and bosons that starts in a metastable vacuum and gradually decays to
the true one. The process considered here will be the nucleation of ``bubbles''
 of true vacuum inside the false one via thermal activation. Our main purpose
will be the analysis of the stability of these ``bubbles'' and the calculation
 of the decay rate as a function of time, in the presence of a finite density
 of fermions at finite temperature. As already shown in reference \cite{CAC1},
 the decay rate is time-dependent as opposed to the time-independent results
previously found in the literature \cite{langer,callan}. This is a direct
consequence of the use of the non-equilibrium formalism. The results presented
 here appear to be in qualitative agreement with  some experimental results.

The introduction of fermions that interact with the scalar field has two
remarkable features. The first, already used in reference \cite{fraga}, is
the preservation of the functional form of the bosonic static solution of the
 equation of motion for the case with no fermions. Through the inverse
scattering methods used here, this fact is rigorously demonstrated. The second
 is the appearance of metastability within metastability. Besides the
metastability which is inherent to our choice of the metastable minimum
 of the effective bosonic potential, we find that the extrema of such a
 potential may be metastable ``bubbles''. Thus, throughout this article,
 the metastable state (for example, the first excited state of a
 {\it cis}-polymer) may develop defects which correspond to either
 unstable or metastable ``bubbles''.

The paper is organized as follows: in section \ref{bosonic}, we briefly
review the results for just a scalar field, which already appeared in the
 literature; in section \ref{inclusion}, we include fermion fields and analyze
 their influence on the effective theory for the bosons; in section
 \ref{chemical}, we add the effects of finite temperature and of a chemical
 potential for the fermions; in section \ref{concl}, we present some comments
 about the method and our conclusions. Some mathematical definitions are left
 for a final Appendix.

\section{The bosonic case: a summary}
\label{bosonic}

In this section we summarize results concerning the decay of metastable
 bosonic systems. Most of the material presented here may be found in
various recent papers \cite{Boy1,CAC1}.

\subsection{The Lagrangian}

Our bosonic system is represented by a real one-dimensional scalar
 field, $\phi(x)$. The dynamics is given by a Lagrangian with the
following form

\begin{equation}
\label{lagrangeana}
{\cal L}={1\over 2}(\partial_\mu\phi)(\partial^\mu\phi)-\lambda
(\phi-\phi_-)^2 \phi(\phi-\phi^*)
\end{equation}

\noindent
where

\begin{equation}
\phi^*\equiv\phi_-\left(1-{m^2\over 2\lambda\phi_-^2}\right)
\end{equation}

The Lagrangian (\ref{lagrangeana}) has an asymmetric double well
potential,$V(\phi)=\lambda
(\phi-\phi_-)^2 \phi(\phi-\phi^*)$, with $V''(\phi)=m^2$, where $m^2$
 is the squared mass of small oscillations around the local minimum
 $\phi_-$ (see Figure 1) and $\lambda$ is a coupling constant.

\subsection{The ``sphaleron'' solution}

The equation of motion associated with (\ref{lagrangeana}) has a static
 solution known as ``sphaleron'' \cite{Manton}. It corresponds to a
 field configuration that starts at the false vacuum $\phi_-$, almost
 reaches the true vacuum $\phi_+$, and returns to $\phi_-$. It looks
 like a ``droplet'' and its explicit form is given by \cite{CAC1,CAC2}

\begin{equation}
\label{sphaleron}
\phi_{sph}(x)=\phi_-+{m\over 2\sqrt{2\lambda}} \left\{\tanh\left[{m\over
2}(x-x_{c.m.}) +s_0\right]-\tanh \left[{m\over
2}(x-x_{c.m.})-s_0\right]\right\}
\end{equation}

\noindent
where

\begin{eqnarray}
s_0&\equiv&{1\over 2}{\rm cosh^{-1}} \left({\varepsilon+1\over
\varepsilon-1}\right)\\
\varepsilon&\equiv&{m^2\over 2\lambda\phi^2_-}
\end{eqnarray}

The parameter $x_{c.m.}$ reflects the translational invariance of the
 equation of motion, $\epsilon$ gives a measure of the difference in
depth between the true and the false vacua and $s_0$ is related to what
 we may call the radius of the ``sphaleron'' (see Figure 2). As we will
 see later, this latter parameter assumes great importance in the
analysis of stability.

It proves convenient to expand the field $\phi(x)$ and and its conjugated
 momentum $\pi(x)$ around the ``sphaleron'' configuration, so as to express
 the Hamiltonian in this particular basis. Collecting terms up to second
 order, the Hamiltonian acquires the following form

\begin{equation}
\label{hamiltoniana}
\hat{\cal H}=E_{sph}+{\hat\pi^2_0\over 2}+{\hat\pi^2_{-1}\over
2}-{\Omega^2\hat\phi^2_{-1}\over 2}+ {1\over 2}\sum_{l\ge 1}(\hat \pi^2_l
+\omega^2_l \hat\phi^2_l)
\end{equation}

\noindent
where $\{\omega_l\}$ is the set of eigenvalues of the fluctuation operator.
 This form of the Hamiltonian clearly isolates the contributions coming from
 the energy of the ``sphaleron'' itself, the collection of harmonic oscilators,
 the translational mode and, most important, the presence of an inverted
oscillator with frequency $\Omega$. This last contribution signals the
existence of an unstable direction in funcional space (associated with
$\phi_{-1}$,which itself is associated with $s_0$ \cite{CAC1}) ready to
guide the decay. Indeed, a short   glance at the behavior of the energy
 of the ``bubble''\footnote{We define here a new structure, that we call
 ``bubble'', that generalizes the ``sphaleron'' in the sense that the
parameter $s_0$ is promoted to the status of a dynamical variable called $s$.}
 as a function of its ``radius'', $s$, shows us three possible regimes: for
 $s<s_0$, the ``bubble'' shrinks and disappears; for $s>s_0$, the ``bubble''
 grows without limit and, for $s=s_0$, we have the critical ``bubble'',
ready to fall (see Figure 3).

\subsection{Time evolution and decay rate}

As we have seen in the last section, the decay of our metastable vacuum
may take place through the nucleation, via thermal activation, of ``bubbles''
 with radius larger than the critical $s_0$. In order to obtain the decay
 rate for this process, we must calculate the probability current along the
 unstable direction, at the saddle point.

Assuming as our initial density matrix one that represents a set of harmonic
 oscillators of frequency $\omega_k$ centered at the metastable minimum
$\phi_-$, and integrating over all the degrees of freedom but $l=-1$, one
obtains, after some algebraic manipulations, the reduced initial density
 matrix \cite{CAC1} (projected onto the unstable direction, $l=-1$, of
 functional space

\begin{equation}
\label{reduzida}
\rho_r(\eta_{-1},\eta'_{-1})={\cal N}\exp\left\{-{1\over 2\hbar}
\left[\alpha (\eta^2_{-1}+\eta^{'2}_{-1})+ 2\gamma \eta_{-1}
\eta'_{-1}\right] \right\}
\end{equation}

\noindent
where

\begin{eqnarray}
{\alpha\over \hbar}&\equiv&(K_1)_{-1,-1}-{1\over 2}\ \vec Q{}^T\tilde
K{}^{-1} \vec Q\\
\noalign{\bigskip}
{\gamma\over \hbar}&\equiv&-(K_2)_{-1,-1}-{1\over 2}\ \vec Q{}^T\tilde
K{}^{-1} \vec Q\\
\noalign{\bigskip}
{\cal N}&\equiv&\sqrt{E_{sph}}\ \left[{\det(K/\pi)\over \det(\tilde
K/\pi)}\right]^{1/2}
\end{eqnarray}

\noindent
and $\eta_l \equiv \phi_l - \bar\phi_l$ ($\phi_--\phi_{sph}(x-x_0)=
\sum_lf_l(x-x_0)\ \bar\phi_l\ ;
\quad \bar\phi_0=0$, see \cite{CAC1}). The kernels are defined in
 the Appendix.

The time evolution of this matrix is dictated by the reduced Hamiltonian

\begin{equation}
\label{Hreduzida}
\hat{\cal H}_{-1}={1\over 2}\left[-\hbar^2\,{\delta^2\over
\delta\phi^2_{-1}} -\Omega^2\phi^2_{-1}\right]
\end{equation}

\noindent
through the Liouville equation. Using the most general gaussian
 {\it ansatz} compatible with unitarity for the time-evolved density
 matrix, one obtains for the decay rate per unit volume (see \cite{CAC1})

\begin{eqnarray}
\label{taxa}
{\Gamma(t)\over L}&=&J[saddle]=J[\eta_{-1}]_{\eta_{-1}=0}=
\nonumber \\
\noalign{\bigskip}
&=&-\Omega\bar\phi_{-1}(0)\ A(t)\ \sqrt{E_{sph}}\ {\cal N}(0)
\exp\left\{ -{1\over \hbar}\, \bar\phi^2_{-1}(0)[{\cal R}(\alpha(0))+\gamma
(0)]\ B(t)\right\}
\end{eqnarray}

\noindent
where

\begin{equation}
\label{coefA}
A(t)\equiv{W^2\over \Omega^2}\ {\senh(\Omega t)\over [\cosh^2(\Omega
t)+{W^2\over \Omega^2}\sinh^2(\Omega t)]^{3/2}}
\end{equation}

\begin{equation}
\label{coefB}
B(t)\equiv{1\over 1+{W^2\over \Omega^2}\ \tanh^2(\Omega t)}
\end{equation}

The form of the decay rate as a function of time is given in Figure 4.
 To understand this plot, we should remember our initial hypothesis.
 We have defined a particular initial state for our system and decided
 to observe its ``natural'' evolution, i.e., an out-of-equilibrium
 evolution which does not replenish the false vacuum. Therefore, the
 insignificant overlap of the initial state with the saddle point
 forces $\Gamma(t)$ to vanish at $t=0$. On the other hand, for $t>0$
 we have the decay to the true vacuum. Thus, $\Gamma(t)$ must vanish
 also for $t \rightarrow \infty$. Further discussions of these
 results can be found in the literature \cite{fraga,Boy1,CAC1}.

\section{Including fermions}
\label{inclusion}

\subsection{Effective action}

The motivation for the inclusion of fermions in our scheme went beyond
 a simple generalization of the formalism. It was based on the belief
 that they would introduce qualitatively different results for the
 stability of the ``bubble'' structures. This hope relied on the Pauli
 exclusion principle. As a consequence of the repulsive interaction
 of the fermions, we expected to find non-trivial metastable minima
 in the function $E=E[s]$.

Throughout this section, we will be interested on the effects of
 fermions on the bosonic field. The natural way to proceed is to
obtain, from an original fermion-boson Lagrangian, an effective
theory for the bosons. The original Lagrangian is defined as

\begin{equation}
\label{lagrangeana2}
{\cal L}={1\over 2}(\partial_\mu\phi)(\partial^\mu\phi)-[V(\phi) -
V(\phi_2)] + \bar\psi_a(i\not\!\!\partial-\mu-g\phi)\ \psi_a
\end{equation}

\noindent
where $\mu$ is the bare mass of the fermions, $g$ is the coupling
 constant, $\psi_a(x)$ is the fermion field, $a$ denotes fermion
 species and $\phi_2$ is a local minimum of the potential
 \footnote{We may find a physical realization of this form
of potential in the description of conducting polymers \cite{yulu}.}

\begin{equation}
\label{potencial2}
V(\phi)={g^2\over 2}(\phi-\phi_0)^2\left(\phi+\phi_0+{2\mu\over
g}\right)^2 + j\phi
\end{equation}

\noindent
where $\phi_0$ is a constant and $j$ is an external current,
 responsible for the asymmetry of the potential even in the
purely bosonic case.

We may now integrate the generating functional

\begin{equation}
\label{gerador}
Z=\int[D\phi][D\psi_a][D\bar\psi_a]\exp\left\{i\int^T_0dt\int^\infty_{-\infty}
dx \ {\cal L}[\phi,\psi_a,\bar\psi_a]\right\}
\end{equation}

\noindent
over the fermions, in order to obtain an effective action
for the bosons. Following the methods of {\it Dashen et al.}
 \cite{dhn}, we arrive at the following expression for the
 effective action

\begin{equation}
\label{acao}
S_{eff}[\phi]=\int^T_0dt\int^\infty_{-\infty}dx\left[{1\over
2}(\partial_\mu\phi)^2- [V(\phi)-V(\phi_2)]\right]+N\sum_i
\alpha_i[\phi] -\sum_in_i\alpha_i[\phi]
\end{equation}

\noindent
where $N$ represent the number of fermion ``species'',
${n_i}$ the occupation number of the discrete (bound states)
 fermionic levels and ${\alpha_i}$ are the Floquet indices,
defined such that

\begin{equation}
\label{Floquet}
\psi(x,t+T)=e^{-i\alpha_i}\,\psi(x,t)
\end{equation}

Assuming the existence of only two symmetric bound states
 and the possibility of ``doping'', which will characterize
 particular configurations defined by the occupation numbers
 $n_{\pm1}$ of the bound states, we may write the effective
 action as

\begin{eqnarray}
S_{eff}[\phi]&=&\int^T_0dt\int^\infty_{-\infty}dx\left[{1\over
2}(\partial_\mu\phi)^2- [V(\phi)-V(\phi_2)]\right]+\nonumber \\
&+&-\int^T_0dt\int^\infty_{-\infty}dx\left[
{g^2\over 2}(\phi^2-\phi^2_2)+b^*g^2(\phi-\phi_2)\right]+\nonumber \\
&+&N\sum_{i<-1}(\alpha_i[\phi]-\alpha_i[\phi_2])-(n_+- n_-)(\alpha_{1} [\phi]-
\alpha_{1} [\phi_2])
\label{acao2}
\end{eqnarray}

\noindent
where we have already introduced counterterms and renormalization
 conditions \cite{CAC3}(that define $b^*$ through
$Z_1(\phi_2,\Lambda,g,\mu,j)=b^*g^2$). Note that $n_{\pm}$
 represents the occupation number of particle states only,
 as opposed to $n_{\pm1}$ which represents particle and
 anti-particle states ($n_+ \equiv n_{+1}; n_- \equiv N-n_{-1}$).

\subsection{SPA approximation}

In order to obtain time-independent solutions, analogous to
the ``sphaleron'', it proves useful to rewrite the action
 (\ref{acao2}) in terms of scattering data. Using well known
 results from inverse scattering methods \cite{CAC3,Novikov,Aoyama},
we may write the effective action as

\begin{eqnarray}
{S_{eff}[\phi]\over T}&=&-{I_3\over
g^2}-2\left[(\phi^2_2-\phi^2_0)+{2\mu\over g}(\phi_2-\phi_0)\right]\,
I_1 -{j\over g}\, I_0-\nonumber\\
&-&(I_1-\mu I_0)-b^*gI_0- {NI_2\over \pi}-(n_+-n_-) [\omega_0 (\phi)-
\omega_0 (\phi_2)]
\label{acao3}
\end{eqnarray}

\noindent
where $\omega_i \equiv \frac{\alpha_i}{T}$ and $I_0$, $I_1$, $I_2$
 and $I_3$ have their explicit forms shown in the Appendix. Now, we
may extremize the action with respect to the scattering data, namely,
 the reflection coefficients $r_{\pm}(k)$ and
 $K^{\pm}_0=(m_F^2-\omega_0^2)^{1/2}$,

\begin{equation}
\label{extremos1}
{\delta S_{eff}\over \delta r_+}={\delta S_{eff}\over \delta r_-}=0
\end{equation}

\begin{equation}
\label{extremos2}
{\delta S_{eff}\over \delta K_0^+}={\delta S_{eff}\over \delta K_0^-}=0
\end{equation}

\noindent
{}From (\ref{extremos1}) we obtain $r_+(k)=r_-(k)=0$, ie,
 the potential is reflectionless. Using this, we may write
 the energy as

\begin{eqnarray}
E&=&m_F\left\{{8\over 3}\left({m_F\over
g}\right)^2\sen^3\theta_0+2\left({N\over \pi}-{\Delta\over m_F}\right)
\sen\theta_0 +N\left(1-{2\theta_0\over \pi}+{\Delta n\over N}\right)
\cos \theta_0\right.-\nonumber\\
\noalign{\bigskip}
\qquad&-&\left.\left({\Delta\over
m_F}\right)\ln\left({1-\sen\theta_0\over 1+\sen\theta_0}\right) -
(N+\Delta n)\right\}
\label{energia}
\end{eqnarray}

\noindent
and the equation for the extrema (the gap equation) as

\begin{equation}
\label{eqextr}
\theta_0+{\pi\over N}\left({\Delta\over m_F}\right)\tan
\theta_0+ {2\pi\over N}\left({m_F\over g}\right)^2 \sen2\theta_0
={\pi\over 2}\left(1+{\Delta n\over N}\right)
\end{equation}

\noindent
where we have introduced the following definitions

\begin{equation}
m_F \equiv \omega(\phi_2)=\mu+g\phi_2=(K_0^2+\omega_0^2)^{1/2}
\end{equation}

\begin{equation}
\Delta n \equiv n_+ - n_-
\end{equation}

\begin{equation}
\omega_0 \equiv m_F\cos \theta_0
\end{equation}

\begin{equation}
\Delta \equiv \mu-gb^*-j/g
\end{equation}

Equation (\ref{eqextr}) determines the possible values for
 $K_0$. From these, together with the Gel'fand-Levitan-Marchenko
 equation \cite{CAC3,Novikov,Aoyama}, we may obtain the explicit
 form of $\phi_{bubble}(x)$

\begin{equation}
\label{campo}
\phi_{bubble}(x)=\phi_2-{K_0^2\over g\omega_0}\,{1\over \senh(2K_0s_0)}\left\{
\tanh\left[
K_0(x+s_0)+\delta \right]-\tanh\left[K_0(x-s_0)+\delta\right]\right\}
\end{equation}

\noindent
where

\begin{equation}
\label{vinculo}
K_0s_0\equiv{\rm tanh^{-1}}\left({m_F-\omega_0\over K_0}\right)
\end{equation}

\begin{equation}
\tanh\,\delta\equiv{K_0(m_F-\omega_0)-C_0\omega_0\over
K_0(m_F-\omega_0) +C_0\omega_0}
\end{equation}

\noindent
and $C_0$ is the normalization constant of the bound states.
 Equation (\ref{campo}) may be rewritten as

\begin{equation}
\label{camponovo}
\phi_{bubble}(x)=\phi_2+\phi_p\{\tanh(\xi+\xi_0)-\tanh(\xi-\xi_0)\}
\end{equation}

\noindent
where

\begin{eqnarray}
\xi&\equiv&K_0x+\delta\\ \label{eq4314}
\xi_0&\equiv&K_0s_0\\ \label{eq4315}
\phi_p&\equiv&-{K^2_0\over g\omega_0 \senh(2K_0s_0)}  \label{eq4316}
\end{eqnarray}

Looking at (\ref{camponovo}) we notice that, in spite of the presence
 of fermions, the functional form of the time-independent solution
 remained unchanged. The effect of the fermions is to rescale the
 parameters of the ``bubble''.

\subsection{Stability of the ``bubble'' solution}

In order to have an asymmetrical potential with a false vacuum
 located at positive $\phi_2$ (see Figure 5), we must adjust the
 parameters such that \cite{CAC3}

\begin{equation}
\label{condicao}
2g\bar\phi_e(\bar\phi_e+a)(\bar\phi_e-a)= \mu-gb^*-j/g=\Delta<0
\end{equation}

\noindent
where $\bar\phi\equiv\phi+\mu/g$, $a\equiv (\bar\phi^{2}_0- 1/2)^{1/2}$ and
$\bar\phi_e$ is an extremum of $V(\bar\phi)$.

We are now able to obtain, numerically, the solutions of the
 gap equation (\ref{eqextr}) and the form of the energy of the
 ``bubble'' as a function of its radius. Assuming $N=2$, we
 can plot these results for the interesting cases
 $\Delta n =-2,-1,0$ (see Figures 6 and 7).

Looking at these figures, we immediately see that our
 expectations are fulfilled in several cases. The reader will
 have noticed that the existence of a metastable ``bubble'' is
 associated with ``doping'': for $\Delta n=-2$ (``ground state'')
 we find just the ``sphaleron''.

If we remember that we have an effective theory for the bosons,
 it should be clear that we are in the same situation as at the
 end of section 2.B. The differences are in the rescaling of the
 parameters of $\phi_{bubble}(x)$ and in the form of $E=E[s]$.
 This latter difference will certainly imply a difference
 between the lifetimes of the ``sphaleron'' (the ``bubble''
 of critical radius) and that of the metastable ``bubble''.
 The decay rate may be calculated in the same manner
(see equation (\ref{taxa})) with $E_{sph}$ replaced with
 $E_{sph}-E_{mb}$, $E{mb}$ being the energy of the metastable
 ``bubble'', whenever appropriate.

We should remark that, for temperatures which are low compared
 with $E_{sph}/k_b$, the metastable ``bubbles'' may have an appreciable
 lifetime. Thus, if we consider small oscillations of these structures
 around the metastable minimum of $E=E[s]$, ie, oscillations in the
 size of the ``bubble'', we may be able to detect their presence through
 their charges \footnote{Natural systems for this kind of experiment are
 doped linear polymer chains \cite{yulu}.}.

\section{Finite temperature and chemical potential effects}
\label{chemical}

A natural way to control the ``doping'' mentioned in the last section
 consists in the introduction of a chemical potential, $\epsilon_F$.
 This may be implemented by adding, to our original Lagrangian, a term
 like \cite{CAC4}

\begin{equation}
\label{qp}
{\cal L}_{ch.pot.}=\epsilon_F\,\bar\psi\,\gamma_0\,\psi
\end{equation}

The net effect of this term is to alter the form of our Floquet
 indices to

\begin{equation}
\label{novofloquet}
\alpha_i=(\omega_i-\epsilon_F)\ T
\end{equation}

We may now study the case of finite temperature by doing the
analytic continuation $iT \rightarrow \beta$, so that

\begin{equation}
\label{newfloquet}
\alpha_i=-i\beta(\omega_i-\epsilon_F)
\end{equation}

Using the results of {\it Dashen et al.} \cite{dhn} for the
 generating functional of the fermions

\begin{equation}
{\cal Z}=\left[\prod^\infty_{i=-\infty}(1+e^{-i\alpha_i})\right]^N\ e^{\t N
\sum^\infty _{i=-\infty}(i\alpha_i/2)}
\end{equation}

\noindent
we may calculate the free energy $F=-{1\over \beta}\ln
{\cal Z}$

\begin{eqnarray}
F&=&-N(\omega_0-m_F)-2N\int^\Lambda_0{dk\over 2\pi}\
(\omega-\epsilon_F)-\nonumber \\
\noalign{\bigskip}
&-&{1\over \beta}\ln \left[\sum_{\{n_{\pm1}\}}C(\{n_i\},N)\ e^{-\beta
(n_{+1} +n_{-1})(\omega_0-\epsilon_F)}\right]-\nonumber\\
\noalign{\bigskip}
&-&{4N\over \beta}\int^\Lambda_0{dk\over 2\pi}\ln
(1+e^{-\beta(\omega-\epsilon_F)} )
\label{eq523}
\end{eqnarray}

\noindent
and, finally, the energy due to the fermions,
 $E_F=\beta\,{\partial F\over \partial\beta}+F$,

\begin{equation}
\label{enfermions}
E_F=(\ran{n_+}-\ran{n_-})(\omega_0-m_F)+{NI_2\over
\pi}-4N\left({\partial \tilde\Delta\over \partial\beta}\right)
\end{equation}

\noindent
where we have introduced the following definitions in
 expression (\ref{enfermions})

(a) $1^{st}$ term: contribution from the discrete levels with

\begin{equation}
\ran{n_i}={\sum_{n_i}C(\{n_i\},N)\ n_i\ e^{-\beta n_i(\omega_0-
\epsilon_F)}\over \sum_{n_i}C(\{n_i\},N)\ e^{-\beta
n_i(\omega_0-\epsilon_F)}}
\end{equation}

(b) $2^{nd}$ term: contribution from the ``Dirac sea''.

(c) $3^{rd}$ term: contribution from the continuum states with

\begin{equation}
\tilde\Delta\equiv 2\beta\int^\Omega_{m_F}{d\omega\over
e^{\beta(\omega- \epsilon_F)}+1} {\rm tan^{-1}}\left({K_0\over \sqrt{\omega^2-
m^2_F}}\right)
\end{equation}

\noindent
where $\Omega$ is the cutoff value of the energy (associated to

 the bandwidth in polymer models).

The energy $E_F$ may be rewritten in terms of the parameter
 $\theta_0$ as follows

\begin{eqnarray}
E_F&=&m_F\left\{(\ran{n_+}-\ran{n_-})(\cos\theta_0-1)+{2N\over \pi}
\left[ \left({\pi\over
2}-\theta_0\right)\cos\theta_0+\right.\right.\nonumber \\
\noalign{\bigskip}
\quad &+& \left.\left.\left(1+{\pi\over Ng^2}-\gamma\right)\sen\theta_0
-{\pi\over
2}\right]\right\} -{4N\over m_F}\ {\partial\tilde\Delta\over \partial\beta}
\label{Ef}
\end{eqnarray}

\noindent
where we have introduced the parameter

\begin{equation}
\gamma\equiv{\pi\over Ng^2}\left({\mu-gb^*-j/g\over
\mu+g\phi_2}\right)= {\pi\over Ng^2} \left({\Delta\over m_F}\right)
\end{equation}

\noindent
which defines the form of the potential. Adding to (\ref{Ef})
 the bosonic contribution for the energy of the ``bubble'' we
obtain, as our total energy,

\begin{eqnarray}
E&\!\!\!\!\!\!=\!\!\!\!\!\!&m_F\left\{{8\over 3}\left({m_F\over
g}\right)^2\sen^3\theta_0+ 2\left( {N\over \pi}-{\Delta\over
m_F}\right)\sen\theta_0+N\left(1-{2\theta_0 \over\pi}+{\ran{\Delta
n}\over N}\right)\right. \cos\theta_0-\nonumber\\
\noalign{\bigskip}
\qquad&-&\left.\left({\Delta\over m_F}\right)\ln\left({1-\sen\theta_0\over
1+\sen \theta_0}\right)-(N+\ran{\Delta n})-{4N\over m_F}\ {\partial
\tilde \Delta\over \partial\beta}\right\}
\label{entotal}
\end{eqnarray}

\noindent
where we have defined $\ran{\Delta n} \equiv \ran{n_+}-\ran{n_-}$,
 and the gap equation

\begin{equation}
\theta_0+\gamma\tan\theta_0+{2\pi\over N}\left({m_F\over g}\right)^2
\sen 2\theta_0={\pi\over 2}\left[1+{\ran{\Delta n}\over N}+{4\over
m_F}\, {\partial\over\partial \theta_0}\,{\partial\over \partial \beta}\
\tilde \Delta\right]
\end{equation}

\noindent
to be compared with (\ref{eqextr}). We are now ready to analyze
 the form of $E=E[s]$ in the limits of high and low temperatures.

In the high temperature limit, $\beta|\Omega- \epsilon_F|\ll 1$.
 Thus, we can approximate the values of $\tilde\Delta$ and
 $\ran{\Delta n}$ by

\begin{equation}
\tilde\Delta\approx\beta\int^\Omega_{m_F}d\omega\
{\rm tan^{-1}}\left({K_0\over \sqrt{\omega^2-m^2_F}}\right)=\beta\,{I_2\over 2}
\end{equation}
\begin{equation}
\ran{\Delta n}\approx-{\beta(\omega_0-\epsilon_F)\over
1-\beta(\omega_0-\epsilon_F)}
\end{equation}

\noindent
in order to obtain, for the total energy,

\begin{eqnarray}
E&=&m_F\left\{{8\over 3}\left({m_F\over g}\right)^2\sen^3\theta_0
+2\left({ N\over \pi}-{\Delta\over
m_F}\right)\sen\theta_0+N\left(1-{2\theta_0\over \pi}-\right.\right.\nonumber\\
\noalign{\bigskip}
\quad&-&\left.{\beta\over N}\,{(\omega_0-\epsilon_F)\over
1-\beta(\omega_0-\epsilon_F)}\right) \cos\theta_0-\left({\Delta\over
m_F}\right) \ln \left({1-\sen\theta_0\over 1+\sen\theta_0}\right)-\nonumber\\
\noalign{\bigskip}
\quad&-&\left(N-{\beta\over N}\,{(\omega_0-\epsilon_F)\over 1-\beta(\omega_0
-\epsilon_F)}\right) -4N\left[\left({\pi\over
2}-\theta_0\right)\right.\cos\theta_0+\nonumber \\
\noalign{\bigskip}
\quad&+&\left.\left.\left(1+{\pi\over
Ng^2}-\gamma\right)\sen\theta_0-{\pi\over 2}\right]\right\}
\end{eqnarray}

\noindent
The form of $E=E[s]$ is shown, for various values of the chemical
 potential, in Figure 8. From this figure, we clearly see that high
 temperatures prevent the appearance of metastable ``bubbles'', just
 yelding dissociation curves.

For low temperatures, we have $\beta|m_F- \epsilon_F|\gg 1$. The
curves for $E=E[s]$ are shown in Figure 9 for the same values of
 $\epsilon_F$\footnote{For $\epsilon_F<0$, we have the same results
 with anti-particles in the place of particles.}. In this case,
temperature plays a small role, as $kT$ is small compared to the
other energy scales. Our results are, thus, analogous to the ones
 we encountered when we considered ``doping'' via a choice of
 occupation numbers.

\section{Conclusion}
\label{concl}

The aim of this work was the study of metastable systems, with
particular interest in the stability of ``bubble'' structures.
The novelty, in comparison with other recent papers, is the presence
 of fermions and the exciting role they play. As we have seen, the
presence of fermions creates a qualitatively different scenario for
 the evolution of the ``bubbles''. Depending almost exclusively on
``doping'' (relative occupation of bound fermionic states), which
may be implemented by means of a chemical potential, we may find
different regimes for the stability of the ``bubbles''. More precisely,
 for non-trivial ``doping'' we encounter a phenomenon of ``quantum
stabilization'' that brings about metastable ``bubbles''. This striking
 feature, that comes as a consequence of the introduction of fermions,
may cause important changes in the physics of the models used to describe
 the systems mentioned in the Introduction.

These results are a great stimulus for generalizing this work to higher
 dimensions. However, there are some points in our scheme that require
 improvement: the inclusion of spinodal decomposition to compete with
 the nucleation mechanism, and a careful study of tunneling, as a competitor
 of thermal activation, together with an improvement of some approximations,
 such as the use of the inverted oscilator \cite{Boy1,CAC1,barton}, are some
 of the points to be refined in order to bring the formalism closer to reality.

\bigskip\bigskip

{\bf Acknowledgements}

E.S.F. would like to thank R.M. Cavalcanti and D. Barci for stimulating
 discussions. E.S.F. and C.A.A.C. would like to thank D. Boyanovsky for
  very fruitful conversations and collaboration. The authors acknowledge
 CNPq for finantial support.

\bigskip\bigskip

\newpage
{\bf Appendix}

(a) Definition of the kernels:

\begin{equation}
K_1(x-y)\equiv\int{dk\over 2\pi}\ e^{-ik(x-y)}\ \left[{\omega_k\
\cosh(\beta \hbar \omega_k)\over \hbar\ \senh(\beta\hbar\omega_k)}\right]
\end{equation}

\begin{equation}
K_2(x-y)\equiv\int{dx\over 2\pi}\ e^{-ik(x-y)}\ \left[{\omega_k
\over \hbar\ \senh(\beta\hbar\omega_k)}\right]
\end{equation}

\begin{equation}
(K_i)_{ll'}\equiv\int dxdy\ {dk\over 2\pi}\ e^{-ik(x-y)} K_i(k)\
f_l(x-x_{c.m.}) \ f_{l'}(y-x_{c.m.})
\end{equation}

\begin{eqnarray}
\tilde K_{ll'}&\equiv&(K_1-K_2)_{ll'}\ ; \ l,l'\ne-1\\
Q_l&=&(K_1-K_2)_{-1,l}\ ; \ l\ne -1
\end{eqnarray}

\noindent
where $\{f_l(x)\}$ are the eigenfunctions of the fluctuation operator.

(b) Definition of the inverse scattering integrals \cite{CAC3}:

\begin{equation}
I_0\equiv{1\over 4\pi i}{\cal P}\int^\infty_{-\infty} dq \left[{P_+(q)\over
im_F+q}+{P_- (q)\over i m_F-q}\right]+\ln\left({m_F-K_0\over
m_F+K_0}\right)
\end{equation}

\begin{equation}
I_1\equiv-{1\over 2}\left\{{1\over 2\pi}\, {\cal P}\int dq \ P_+(q) +2K^+_0+
[(+)\leftrightarrow (-)]\right\}
\end{equation}

\begin{equation}
I_2\equiv{1\over 2} \left\{\int^\Lambda_0{kdk\over \sqrt{k^2+m^2_F}}
\left({\cal P} \int^\infty_{-\infty}dq\,{P_+(q)\over k-q}+2 {\rm arctan}
\left({K^+_0\over k}\right)+[(+)\leftrightarrow (-)]\right)\right\}
\end{equation}

\begin{equation}
I_3\equiv-{1\over 2}\left\{ {2\over \pi} {\cal P}\int^\infty_{-\infty} dq
\ q^2 \ P_+(q)-{8\over 3}(K^+_0)^3+[(+)\leftrightarrow (-)]\right\}
\end{equation}
\bigskip
\noindent where ${\cal P}$ means principal value, $K^+_0=K^-_0$ and
$P_\pm(q)\equiv\ln(1-|r_\pm(q)|^2)$.

\newpage

\underline{\bf Figure Captions:}

\vspace{3mm}

{\bf Figure 1}: Form of the potential.

\vspace{3mm}

{\bf Figure 2}: The ``sphaleron'' solution.

\vspace{3mm}

{\bf Figure 3}: ``Bubble'' energy as a function of its radius.

\vspace{3mm}

{\bf Figure 4}: Decay rate as a function of time.

\vspace{3mm}

{\bf Figure 5}: Form of $[V(\phi)-V(\phi_2)]$.

\vspace{3mm}

{\bf Figure 6}: Numerical solution of (\ref{eqextr}) for (a)$\Delta n=0$,
 (b)$\Delta n=-1$ and (c)$\Delta n=-2$.

\vspace{3mm}

{\bf Figure 7}: Energy of the ``bubble'' as a function of its radius for
 (a)$\Delta n=0$, (b)$\Delta n=-1$ and (c)$\Delta n=-2$. Equation
(\ref{vinculo})
 imposes the restriction $s_0 \geq 1/2m_F$.

\vspace{3mm}

{\bf Figure 8}: Energy of the ``bubble'' as a function of its radius
 in the limit of high temperature.(a) $m_F<\epsilon_F
<\Omega$. (b) $\omega_0<\epsilon_F<m_F$. (c) $0<\epsilon_F<\omega_0$.

\vspace{3mm}

{\bf Figure 9}: Energy of the ``bubble'' as a function of its radius
 in the limit of low temperature.(a) $m_F<\epsilon_F
<\Omega$. (b) $\omega_0<\epsilon_F<m_F$. (c) $0<\epsilon_F<\omega_0$.

\end{document}